\documentclass{aa}  

\usepackage{graphicx}
\usepackage{txfonts}
\begin{document}

   \title{Reevaluating the cosmological redshift: insights into
  inhomogeneities and irreversible processes}

\date{\today}

   \author{P. Tremblin
          \inst{1,2}
          \and
          G. Chabrier\inst{3,4}
          }

   \institute{Université Paris-Saclay, UVSQ, CNRS, CEA, Maison de la Simulation,
     91191, Gif-sur-Yvette, France
     \and
     Université Paris-Saclay, Université Paris Cité, CEA, CNRS, AIM, 91191
     Gif-sur-Yvette France\\
      \email{pascal.tremblin@cea.fr}
     \and
     Ecole Normale Supérieure de Lyon, CRAL, UMR CNRS 5574, 69364 Lyon
     Cedex 07, France
     \and
     Astrophysics Group, University of Exeter, EX4 4QL Exeter, UK\\
      \email{chabrier@ens-lyon.fr}
             }

 
  \abstract
      {}
   { Understanding the expansion of the Universe remains a
   profound challenge in fundamental physics. The complexity of solving General
   Relativity equations in the presence of intricate, inhomogeneous flows has
   compelled cosmological models to rely on perturbation theory in a
   homogeneous FLRW background. This approach accounts for a redshift of light
   encompassing contributions from both the cosmological background expansion along the
   photon's trajectory and Doppler effects at emission due to
   peculiar motions. However, this computation of the redshift is not covariant,
   as it hinges on specific coordinate choices that may distort physical
   interpretations of the relativity of motion.}
   {In this study, we show that peculiar motions, when tracing
the dynamics along time-like geodesics, must contribute to the redshift of light
through a local volume expansion factor, in addition to the background expansion. By
employing a covariant approach to redshift calculation, we address the central
question of whether the cosmological principle alone guarantees that the
averaged local volume expansion factor matches the background expansion.}
   { We
establish that this holds true only in scenarii characterized by a reversible
evolution of the Universe, where inhomogeneous expansion and compression modes mutually
compensate. In the presence of irreversible processes, such as the dissipation
of large-scale compression modes through matter virialization and associated
entropy production, the averaged expansion factor becomes dominated by expansion
in voids that cannot be compensated anymore by compression in virialized
structures. Furthermore, a Universe where a substantial portion of its mass has
undergone virialization, adhering to the background evolution on average leads
to significant violations of the second law of thermodynamics. Our approach
shows that entropy production due to irreversible processes during the formation
of structures plays the same role as an effective, time-dependent cosmological
constant, i.e. dynamical dark energy, without the need to invoke new unknown
physics. Our
findings underscore the imperative need to reevaluate the influence of
inhomogeneities and irreversible processes on cosmological models, shedding new
light on the intricate dynamics of our Universe.}
   {}

   \keywords{
               }

   \maketitle
%

\section{Introduction}

The comprehension of the recessional velocity of extragalactic objects hinges
upon the redshift-distance relation derived from homogeneous solutions of
Einstein's field equations within the framework of the
Friedmann-Lemaître-Robertson-Walker (FLRW) metric. Within this coordinate
system, the spatial coordinates are comoving with the matter that fills the
Universe, attributing the cosmological redshift to the expansion of ``space''
along the path of photons. However, observations in the nearby Universe have
revealed a notable scatter around the linear redshift-distance relation,
primarily driven by the presence of peculiar velocities of sources, inducing a
Doppler shift upon photon emission \citep{peebles1993, peacock1998}. The debate in the literature regarding
whether cosmological and Doppler shifts are equivalent phenomena is ongoing
(\cite{bunn2009}, see also \cite{peacock1998, whiting2004, chodorowski2007, peacock2008}). As
noted in \cite{bunn2009}, the interpretation of space
expansion is not entirely satisfactory from a relativistic standpoint, as it
relies on a specific choice of coordinates, lacking
covariance. Conversely, \cite{rasanen2009} introduced a covariant approach to
redshift calculation for an inhomogeneous dust universe. While this approach
represents a significant advancement in understanding the nature of redshift in
a manner consistent with the covariant framework of General Relativity, it has
been relatively overlooked. This is largely because, on average, the results do
not seem to deviate from the background FLRW expansion in a statistically homogeneous
and isotropic universe \citep{rasanen2009, rasanen2010}.

This paper aims to reevaluate the cosmological redshift by addressing the
influence of inhomogeneities and irreversible processes. In
Sect.~\ref{sect:non-comoving}, we begin by providing a
detailed analysis on the standard interpretation of the redshift within the
FLRW metric and discuss the limitations of this approach by using non-comoving
coordinates. In the subsequent section (Sect.~\ref{sect:cov}), building on
\cite{rasanen2009}, we introduce a covariant approach to the redshift 
calculation, emphasizing the role of peculiar motions and local volume expansion
factors. We propose a new approach to the redshift calculation by introducing a
tracer of time-like geodesics, the cosmological redshift is associated to the
local expansion rate along these time-like geodesics while the Doppler shifts
are associated to the potential non-geodesic motions of the source and
observer. Following this, we present in Sect.~\ref{sect:irr} our findings on the
impact of irreversible processes
on the cosmological redshift. We argue that in scenarios characterized by
irreversible evolution, such as the dissipation of large-scale compression modes
through matter virialization and associated entropy production, the averaged
expansion factor is dominated by expansion in voids. This leads to significant
violations of the second law of thermodynamics if the Universe's evolution
adheres to the background evolution on average. Lastly, we discuss in
Sect.~\ref{sect:obs} the broader implications of our findings for
cosmological models and observations. We conclude in Sect.~\ref{sect:con} on the
necessity of reevaluating the influence
of inhomogeneities and irreversible processes to better understand the complex
dynamics of the Universe.

\section{Redshift in a homogeneous expanding universe in non-comoving coordinates}\label{sect:non-comoving}

We begin with a standard flat
FLRW metric with a scale factor, $a(t)$, such that
\begin{equation}
 ds^2 = -c^2 dt^2 + a^2(dr^2 + r^2 d\Omega^2).
\end{equation}
The proper distance $R$ is
\begin{equation}
R = a(t) r,
\end{equation}
and the total velocity is then
\begin{equation}
\frac{dR}{dt} = H(t) R + a(t) \frac{dr}{dt},
\end{equation}
with $H\equiv \dot{a}/a$. Within the total velocity, one can discern a
contribution originating from the recession velocity due to the expansion of
space, and another
contribution arising from peculiar velocities $v_p = a(t) dr/dt$.
In the late evolution of the Universe, peculiar motions are non-relativistic in the
weak field limit.
Perturbation theory then leads to the following formulation of the
redshift with a cosmological contribution from space expansion and a contribution from
the Doppler shift associated to peculiar motions
\begin{equation}\label{eq:pert-redshift}
1+z \approx \exp\left(\int_{t_\mathrm{src}}^{t_\mathrm{dst}} H dt\right)\left(
1+ \frac{v_p}{c}\right),
\end{equation}
assuming for simplicity that the observer at time $t_\mathrm{dst}$  has no
  peculiar motion, and the
  source at time $t_\mathrm{src}$ a peculiar motion $v_p$ at emission.
  With this approach, the inhomogeneities have no significant impact on
    the redshift, provided that the peculiar velocities remain small
  $v_p\ll c$.

We now assume a perfectly homogeneous FLRW Universe with no peculiar motion
$v_p=0$. We wish to apply a local coordinate transformation in a non-comoving frame,
characterized by an expansion factor $a_s(t)$ different from $a(t)$ such that 
\begin{eqnarray}
  r_d &=& r a_d(t),\cr
  t_d &=& t + \alpha(r,t),
\end{eqnarray}
with $a_s = a/a_d$ and $\alpha(r,t)$ a function that will be specified
later, such that $t_d \approx t$ at leading order in $O(\dot{a}_dr/c)$. We use
such a coordinate
transformation, locally, such that $\dot{a}_dr/c$ is indeed much smaller than
unity. With this transformation, the proper
distance $R$ is
\begin{equation}
R = a(t) r = a_s(t)r_d,
\end{equation}
with $a_s=a/a_d$. The total velocity is then
\begin{equation}
\frac{dR}{dt} = H_s(t) R + a_s(t) \frac{dr_d}{dt_d},
\end{equation}
with $H_s\equiv \dot{a_s}/a_s$. Within the total velocity, one can discern now a
contribution originating from the recession velocity due to the expansion of
``space'', characterized by the scale factor $a_s$, and another
contribution arising from non-comoving velocities $v_{p,r} = a_sH_dr_d$.

We choose the function $\alpha(r,t)$ such that the metric in $(r_d, t_d)$
coordinates has only perturbative deviations from a homogeneous metric with a scale factor
$a_s$. The coordinate transformation verifies
\begin{eqnarray}
  dr_d &=& a_d dr + r \dot{a}_d dt,\cr
  dt_d &=& (1 + \dot{\alpha})dt + \alpha^\prime dr,
\end{eqnarray}
with $\dot{\alpha}$ the derivative with respect to $t$ and $\alpha^\prime$ the
derivative with respect to $r$. This system can be inverted to give
\begin{eqnarray}
  a_d dr &=& ((1+\dot{\alpha}) dr_d - r \dot{a}_d
  dt_d)/(1+\dot{\alpha}-r H_d\alpha^\prime),\cr
  dt &=& (dt_d - \alpha^\prime /a_d dr_d)/(1+\dot{\alpha}-r H_d\alpha^\prime),
\end{eqnarray}
with $H_d=\dot{a_d}/a_d$. By injecting these equations into $ds^2$, the
non-diagonal metric coefficient is
\begin{equation}
  g_{r_d t_d} = 2(c^2\alpha^\prime/a_d-a_s^2(1+\dot{\alpha})r\dot{a}_d)/(1+\dot{\alpha}-rH_d\alpha^\prime)^2
\end{equation}
We impose
\begin{equation}
  \alpha(r,t) = \frac{1}{2}\frac{a^2 H_d r^2}{c^2},
\end{equation}
such that the metric in $(r_d, t_d)$ coordinates is given by
\begin{equation}
 ds^2 = -c^2 A dt_d^2 + a_s^2(B dr_d^2 + r_d^2 d\Omega^2) - 2 c dt_d a_s dr_d C ,
\end{equation}
with,
\begin{eqnarray}
  A &=& (1-a_s^2H_d^2r_d^2/c^2)/(1+\dot{\alpha}- a_s^2 H_d^2r_d^2/c^2)^2,\cr
  B &=& ((1+\dot{\alpha})^2-a_s^2 H_d^2 r_d^2/c^2)/(1+\dot{\alpha}- a_s^2
  H_d^2r_d^2/c^2)^2, \cr
  C &=& \dot{\alpha} a_s H_d r_d/c /(1+\dot{\alpha}-a_s^2H_d^2r_d^2/c^2)^2
\end{eqnarray}
At leading order in $O(a_s H_dr_d/c)$, we get
\begin{eqnarray}
  \alpha &=& O(a_s^2 H_d r_d^2/c^2)\cr
  \dot{\alpha} &=& O(a_s^2 H_d^2 r_d^2/c^2)\cr
  \alpha^\prime &=& O(a_s^2 H_dr_d/c^2)\cr
  A &=& 1 + O(a_s^2H_d^2 r^2/c^2) \cr
  B &=& 1 + O(a_s^2H_d^2 r^2/c^2) \cr
  C &=& O(a_s^3 H_d^3 r_d^3/c^3)
\end{eqnarray}
The metric in $(r_d, t_d)$ coordinates has therefore only perturbative
deviations from a homogeneous metric with a scale factor
$a_s$. In this coordinate system and using the same
computation of the redshift as usually done for weak perturbations around
a homogeneous metric, at the leading order in $O(a_s H_d
r_d/c)$, the redshift of light emitted by a source at $r_d = a_d
r_\mathrm{src}, t_d = t_\mathrm{src}$ toward an
observer at $r_d = 0, t_d = t_\mathrm{dst}$ can be expressed as follows:
\begin{equation}
1+z \approx \exp\left(\int_{t_\mathrm{src}}^{t_\mathrm{dst}} H_s dt\right)\left(
1+ \frac{H_d a_s a_d r_\mathrm{src}}{c}\right),
\end{equation}
with two contributions: one stemming from the ``space'' expansion along the
photon path of the homogeneous metric with scale factor $a_s(t)$ and another
originating from a Doppler term from non-comoving
velocities only at emission. At leading order in $O(H_sr/c)$, this is equivalent to
$z\approx (H_s+H_d)a_s a_d r_\mathrm{src}/c = H a r_\mathrm{src}/c$ only when
$H_S$ is independent of time. This method of calculating redshift is not
covariant, as it essentially converts part of the cosmological redshift
along the photon's path into a Doppler shift that occurs solely at the point of
emission.

\section{Covariant computation of the redshift in an inhomogeneous universe}\label{sect:cov}

Following \cite{rasanen2009}, let us begin by defining a set of observers who
are tracing time-like geodesics,
characterized by their four-velocity denoted as $u^\alpha$. Consequently, these
observers satisfy the conditions $u^\beta \nabla_\beta u^\alpha = 0$ and
$u^\alpha u_\alpha = -1$. It is important to note that these observers may not
necessarily move in tandem with the matter filling the Universe, as this matter
can experience acceleration due to non-gravitational forces and may not strictly
adhere to time-like geodesics. However, especially on large scales and within
the context of dark matter, we can consider the dark matter fluid as a useful
tracer of these time-like geodesics.

To further elucidate, we introduce the projection tensor, denoted as $h_{\alpha
  \beta}$, which operates within the tangent space orthogonal to $u^\alpha$. It
is defined as $h_{\alpha \beta}\equiv g_{\alpha \beta}+u_\alpha u_\beta$. With
this tensor, we can proceed to decompose the covariant derivative of
$u^\alpha$
\begin{equation}
\nabla_\beta u_\alpha = \frac{1}{3}h_{\alpha\beta}\theta + \sigma_{\alpha \beta}
+ \omega_{\alpha\beta}
\end{equation}
with $\theta = \nabla_\alpha u^\alpha$, the volume expansion rate,
$\sigma_{\alpha\beta} = \nabla_{(\alpha}u_{\beta)}-h_{\alpha\beta}\theta/3$ the
shear tensor, and $\omega_{\alpha\beta}=\nabla_{[\beta}u_{\alpha]}$, the
vorticity tensor. For simplicity, we will assume in the rest of this study that
the shear and vorticity tensors can be neglected.

We now define $k^\alpha$, the tangent vector of the null geodesics that
satisfies $k^\beta \nabla_\beta k^\alpha = 0$ and $k^\alpha k_\alpha =
0$. The tangent vectors $u^\alpha$ and
$k^\alpha$ are parallel propagated with respect to the time-like and null
geodesics, respectively, but not with respect to each other. Consequently, the
photon momentum changes along the time-like geodesics and the redshift is defined by
\begin{equation}
1+z = \frac{E_\mathrm{src}}{E_\mathrm{dst}}
\end{equation}
with $E_\mathrm{src}$ the photon energy at emission by the source and
$E_\mathrm{dst}$ the photon energy at the location of the observer. The energy
can be computed from $E=-u_\alpha k^\alpha$ and following \cite{rasanen2009} we
decompose $k^\alpha$ into a component parallel and a component orthogonal to the
time-like geodesics $k^\alpha = E (u^\alpha + e^\alpha)$ with $u_\alpha
e^\alpha=0$ and $e_\alpha e^\alpha=1$ (and consequently $h_{\alpha\beta}e^\alpha
e^\beta = 1$). The evolution of the energy along the
null geodesic can then be followed with the affine parameter $\lambda$,
\begin{eqnarray}
  \partial_\lambda E &\equiv& k^\alpha \nabla_\alpha E\cr
  &=& - k^\alpha k^\beta \nabla_\alpha u_\beta \cr
  &=& - E^2 e^\alpha e^\beta \nabla_\alpha u_\beta \cr
  &=& - \frac{E^2}{3}\theta
\end{eqnarray}
  Without vorticity, the hypersurfaces of constant proper time are orthogonal to
  $u^\alpha$, and $t(\lambda)$ is monotonic. We can then invert the relation
  between $\lambda$ and $t$ to obtain $d\lambda = dt/E$, hence
\begin{equation}\label{eq:cov-redshift2}
1+z = \exp\left(\int_{t_\mathrm{src}}^{t_\mathrm{dst}}\frac{\theta(t,\vec{x}(t))}{3} dt \right)
\end{equation}
A more detailed demonstration including vorticity and shear can be found in
\cite{rasanen2009}. The main difference here is that we take a set of observers
following time-like geodesics that are not necessarily co-moving with matter
filling the universe. For a homogeneous metric with scale factor $\tilde{a}(t)$,
 $\tilde{H}(t) = \dot{\tilde{a}}/\tilde{a}$, and for small
three-velocity $\vec{\tilde{u}_g}$, the volume expansion factor is given by $\theta\approx
3\tilde{H}+\vec{\nabla}\cdot \vec{\tilde{u}_g}$. Applying
eq.~\ref{eq:cov-redshift2} to the non-comoving coordinates defined above, we get
\begin{eqnarray}
1+z &\approx&
\exp\left(\int_{t_\mathrm{src}}^{t_\mathrm{dst}}\frac{3H_s+\vec{\nabla}\cdot
  (dr_d/dt_d \vec{e}_{r_d})}{3} dt \right) \cr
  &\approx& \exp\left(\int_{t_\mathrm{src}}^{t_\mathrm{dst}}(H_s+H_d)dt \right).
\end{eqnarray}

With this covariant approach, ``space-time'' geometry characterized by the
expansion rate $H_s$ and non-comoving velocities of observers following
time-like geodesics characterized by $H_d r_d$ contribute both as an expansion
factor along the trajectory of the photons. For a homogeneous FLRW solution with
a Hubble expansion rate $H$,
the comoving coordinate system would lead to $\tilde{H} = H$ and
$\vec{\tilde{u_g}}=0$ and the non-comoving coordinate system defined in
Sect.~\ref{sect:non-comoving} to $\tilde{H} = H_s$ 
and $\vec{\tilde{u_g}}= dr_d/dt_d \vec{e}_{r_d}$. In both coordinate systems, a
consistent calculation of the cosmological redshift is obtained with
$\theta=3H$.

As a simple example, let us consider a universe with $G=0$ such that
   $H=0$ and $a(t)=1$. In such a universe
  the cosmological redshift should be exactly zero, since there is no expansion along
  the photon path and there should be only a Doppler shift from the
  velocity difference between the source and the observer. Let us assume for
  simplicity that they are both at rest such that the total
  redhift is zero. Let us also consider an
  arbitrary coordinate system with a scale factor $a_s(t)$ and an
  expansion rate $H_s(t)$, we have thus $a_d(t)=1/a_s(t)$ and
  $H_d(t)=-H_s(t)$. In such a coordinate system, the perturbative
  redshift calculation gives
  \begin{eqnarray}
1+z& \approx \exp\left(\int_{t_\mathrm{src}}^{t_\mathrm{dst}} H_s(t) dt\right)\left(
1 - \frac{H_s(t_\mathrm{src})  r_\mathrm{src}}{c}\right),\cr
1+z& \approx \left(1 + \int_{t_\mathrm{src}}^{t_\mathrm{dst}}( H_s(t)- H_s(t_\mathrm{src}))dt\right)
  \end{eqnarray}
  which demonstrates that the Doppler shift at emission cannot compensate the
  expansion of the coordinate system along the photon path as soon as this
  expansion is time dependent. However, with the covariant calculation, the
  redshift is exactly zero since $\theta = 3H_s(t) -
  \vec{\nabla}(H_s(t)\vec{r}_d) =0$. This further demonstrates that the
  perturbative formulation of the redshift with the Hubble expansion rate $H$
  instead of the local expansion rate $\theta$ along time-like geodesics
  is not covariant.


Lastly, it is imperative to highlight the precise physical entity that should be traced by the local
volume expansion rate and $\vec{\tilde{u_g}}$. It is crucial to note that the local
volume expansion rate should not mirror the fluid velocity but rather correspond
to the velocity field of a hypothetical set of free-falling observers following
time-like geodesics  \citep{rasanen2009}. The matter filling the universe can
only serve as a tracer
of these time-like geodesics if the acceleration due to non-gravitational forces
can be deemed negligible. Hence, we advocate for the use of expressions that
avoid the term ''expansion of space'', as it lacks covariance, and instead, we
will refer to the expansion along time-like geodesics when discussing
cosmological redshift. Then, the Doppler shifts can be defined as originating
from the potential non-geodesic motion of the source and observer at emission
and reception.

In a inhomogeneous universe with an arbitrary
  expanding coordinate system defined by $\tilde{a}(t), \tilde{H}(t)$, the
  general expression of the redshift can therefore be defined as follows
\begin{eqnarray}\label{eq:cov-redshift-full}
1+z =&
\exp\left(\int_{t_\mathrm{src}}^{t_\mathrm{dst}}\frac{\theta(t,\vec{x}(t))}{3}
dt \right) \times \cr
&\left(
1+ \frac{v_\mathrm{Doppler}(t_\mathrm{src},\vec{x}(t_\mathrm{src}))}{c}
-  \frac{v_\mathrm{Doppler}(t_\mathrm{dst},\vec{x}(t_\mathrm{dst}))}{c}\right)
\end{eqnarray}
with
\begin{eqnarray}\label{eq:cov-redshift-full-def}
  &\theta(t,\vec{x}) = 3\tilde{H}(t) + \vec{\nabla}\cdot \vec{\tilde{u}}_\mathrm{g},
  \cr
  &v_\mathrm{Doppler}(t,\vec{x}) = \tilde{a}(t)
  (\vec{\tilde{u}}_\mathrm{fluid}-\vec{\tilde{u}}_\mathrm{g})\cdot
  \vec{n}.
\end{eqnarray}
Here, we define the cosmological redshift as the integral along the
  line of sight of the local expansion rate along time-like geodesics, and the
  Doppler shifts as the Doppler terms from
  non-geodesic motions of the source and the observer in the non-relativistic
  limit. These definitions are now fully covariant as $\theta$ and the velocity
  difference in the Doppler shift are invariant under a change of the coordinate
  system. Fundamentally, one can see that the cosmological contribution
  should not transform into a Doppler one or vice-versa under a change of the
  coordinate system.
  We consider only the expansion rate in Eq.~\ref{eq:cov-redshift-full}
  since vorticity does not contribute to the redshift and shear is negligible on
  large scale if structures have no preferred orientation \citep{rasanen2009}.

At large
scale and especially with dark matter, the matter within the Universe is
free-falling and
can be taken as a tracer of the dynamics along time-like geodesics. However, in
the cosmological standard model, the coordinate system is
comoving with the FLRW background and is not comoving with these time-like
geodesics.  Consequently, the conventional redshift calculation overlooks the
inhomogeneous dynamics along these geodesics, attributing a Doppler shift solely to
peculiar motions at the point of emission, mirroring our example using
non-comoving coordinates within a homogeneous FLRW universe. In contrast, the covariant
calculation of the redshift highlights that peculiar
motions responsible for the formation of cosmic structures do indeed contribute
to the redshift along the
photon path through the local expansion rate.

This insight leads us to reconsider the influence of inhomogeneities on the
cosmological redshift. Back-reaction is not a question of small
peculiar velocities relative to the speed of light, but requires a deeper inquiry:
whether, in a statistically homogeneous and isotropic universe, the averaged
local volume expansion rate along time-like geodesics provides a close
approximation to the background expansion $3H$, which might not be the case even
if the Doppler shifts of the source and observer relative to time-like geodesics
are negligible.

\section{Impact of irreversible processes on the cosmological redshift}\label{sect:irr}

\begin{figure}[!ht]
\centering
\includegraphics[width=0.7\linewidth]{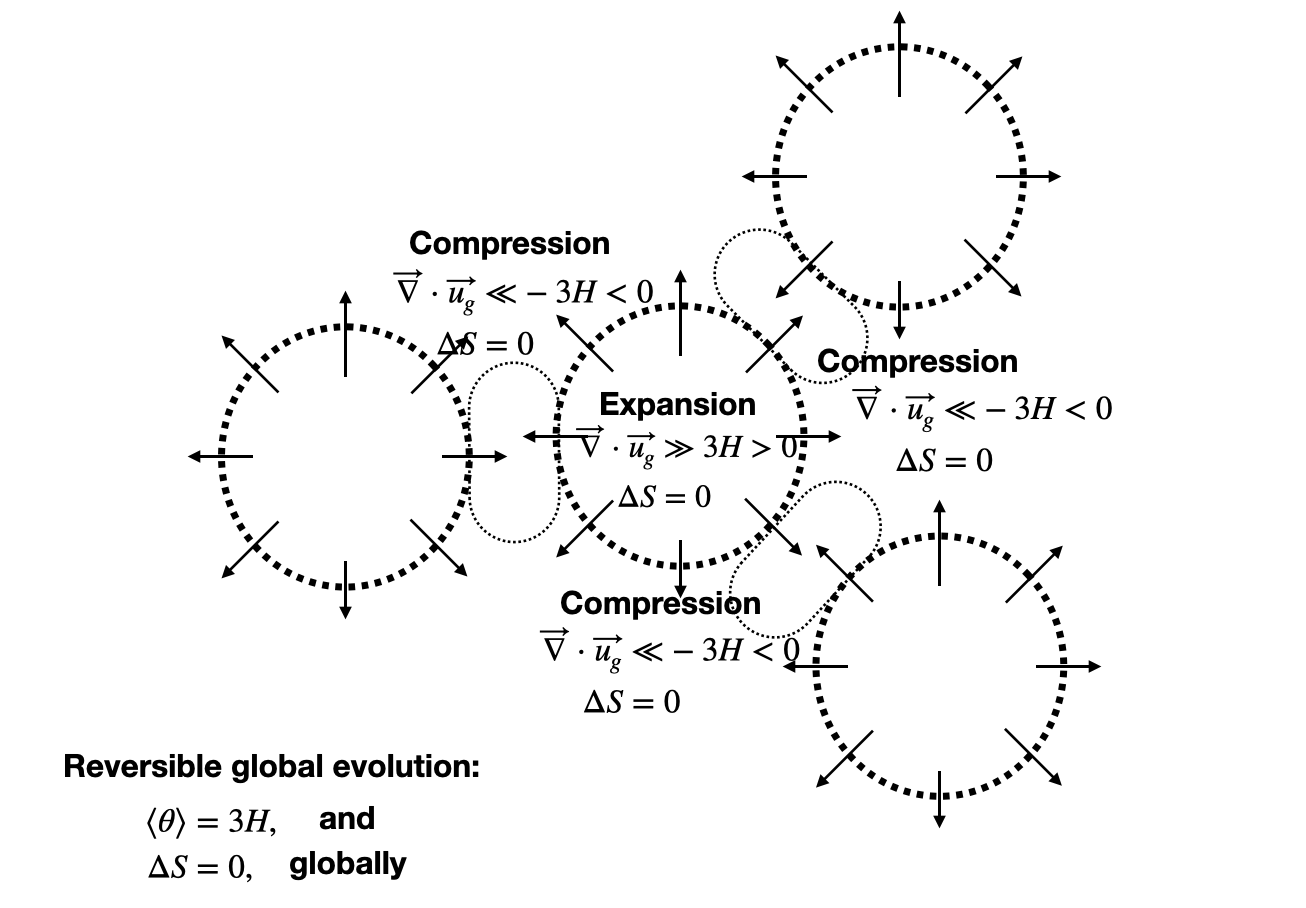}
\includegraphics[width=0.8\linewidth]{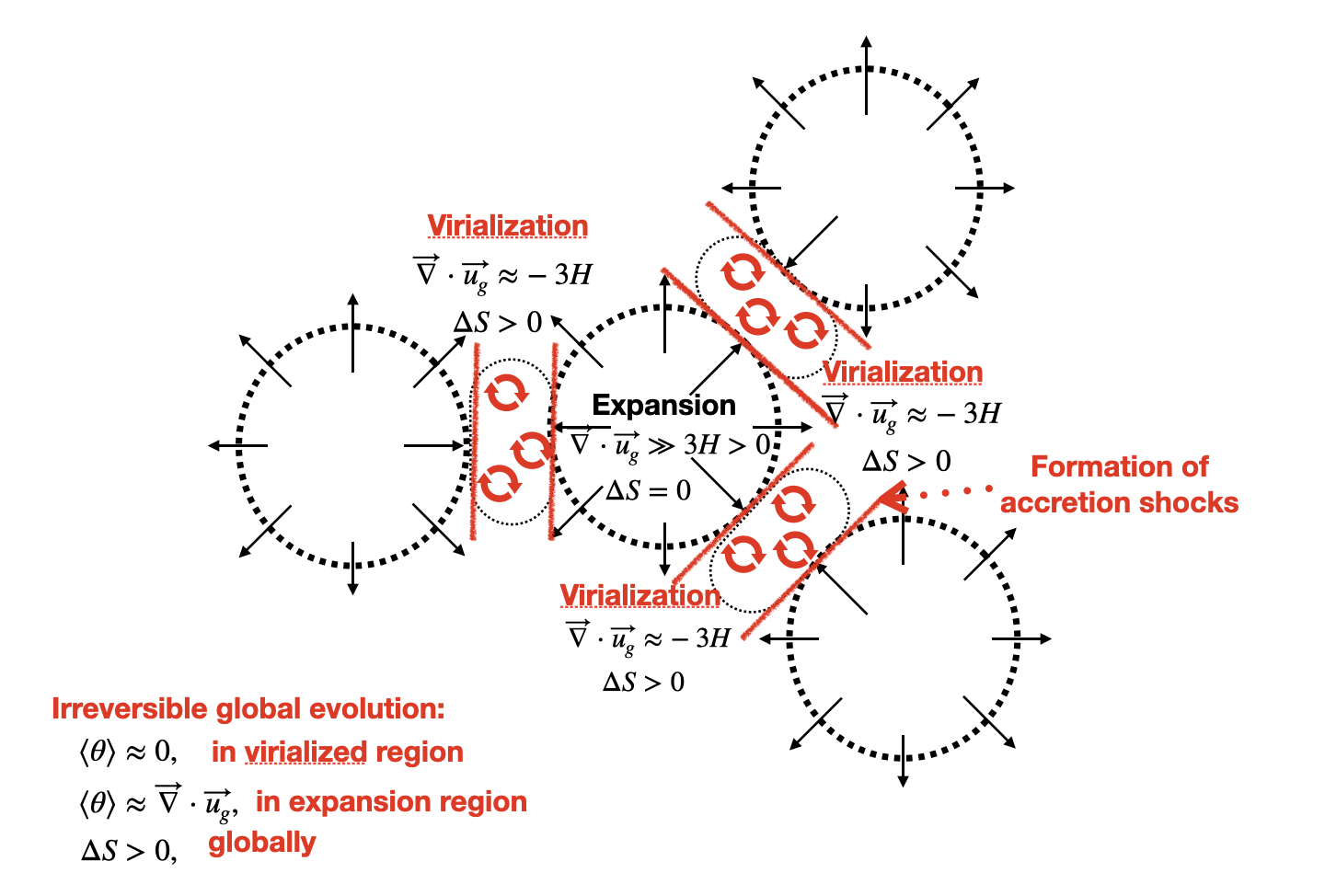}
\caption{Top: reversible evolution with expansion and compression modes. Down:
  irreversible evolution when compression modes are dissipated by virialization
  through accretion shocks.}\label{fig:ir-reversible}
\end{figure}

The condition $\langle \theta \rangle \approx 3H$ can be expressed equivalently as $\langle
\vec{\nabla}\cdot \vec{u_g} \rangle \approx 0$. When we start from a
homogeneous universe where $\vec{u_g}=0$ and consider a reversible
evolution, one might naturally expect e.g. based on hydrodynamics equations, that the condition $\langle
\vec{\nabla}\cdot \vec{u_g} \rangle \approx 0$ would hold . Essentially this implies
that inhomogeneous expansion and compression modes evolve smoothly and mutually compensate on
average, resulting in $\langle \theta \rangle \approx 3H$ (see top panel of
Fig.~\ref{fig:ir-reversible}). It is a common belief that
the cosmological principle i.e. homogeneity and isotropy of the Universe at
large scales, is sufficient to ensure such a condition. In previous
investigations addressing the back-reaction problem
\citep{buchert1997,buchert2008}, which examines the impact
of inhomogeneities on the cosmological redshift, it has been generally assumed that
this condition holds \citep{buchert1997,rasanen2009}. Consequently,
it would appear that back-reaction is negligible in the non-relativistic
limit \citep{ishibashi2006} (but see also \cite{buchert2015}). However, we aim to reevaluate
this conclusion since structure formation does not have inhomogeneous expansion
  and compression modes that mutually compensate on average: compression modes
  are dissipated by virialization and the associated entropy production and
  cannot compensate the expansion of the voids. Hence in
  the presence of irreversible processes, $\langle \theta \rangle \neq 3H$. The
  link between expansion and entropy production can be made explicit by using
  the second principle of thermodynamics assuming that a quasi-equilibrium
  defined by a temperature $T$ and a stress tensor with pressure $\sigma$ has
  been reached by virialization,
  \begin{eqnarray}
    &T dS= dE + \sigma dV,\cr
    &T ds= de + Z k_b T d(\log V),
  \end{eqnarray}
  for a closed system of N particles (dN=0) with an entropy per particle $s=S/N$,
  an internal energy per particle $e=E/N$, and a compressibility factor
  $Z=\sigma V/Nk_bT$.
  Assuming the evolution of an irreversible system from an initial
  state and the evolution of a reversible
  system that reaches the same internal energy $e$, temperature $T$,
  compressibility factor $Z$, and has the same entropy as the initial state,
\begin{eqnarray}\label{eq:redhisft-entropy}
    T \Delta s &=\Delta e + Z k_b T \log
    (V_\mathrm{irrev}/V_\mathrm{ini}),\cr
         0 &=\Delta e + Z k_b T \log (V_\mathrm{rev}/V_\mathrm{ini}),
  \end{eqnarray}
implies that
\begin{eqnarray}
  \frac{V_\mathrm{irrev}}{V_\mathrm{ini}} =
  \frac{V_\mathrm{rev}}{V_\mathrm{ini}}\exp\left(\frac{\Delta s}{k_b Z}\right).
  \end{eqnarray}
We recall that the expansion rate can be directly linked to the variation of
volume by using the equation of mass conservation with a Lagrangian time derivative
\begin{eqnarray}
\frac{d \log V}{dt} = \theta,
\end{eqnarray}
which implies the following expression of the redshift for motions that follow
time-like geodesics in an irreversible system (assuming spatial ergodicity along
the photon path),
\begin{eqnarray}\label{eq:redhisft-entropy}
  1+z &&= \left(\frac{V_\mathrm{irrev}(t_\mathrm{dst})}{V_\mathrm{ini}(t_\mathrm{src})}\right)^{1/3},\cr
  && =
  \left(\frac{V_\mathrm{rev}}{V_\mathrm{ini}}\right)^{1/3}\exp\left(\frac{\Delta s}{3k_b
      Z}\right), \cr
  &&= \frac{a(t_\mathrm{dst})}{a(t_\mathrm{src})}\exp\left(\frac{\Delta s}{3k_b Z}\right)
\end{eqnarray}
  where $a(t)$ is the scale factor of the reversible background
  evolution. Eq.~\ref{eq:redhisft-entropy} demonstrates the direct link between
  entropy production and the cosmological redshift in an irreversible
  system. The second principle of thermodynamics states that the entropy
  production is always positive with $\Delta s>0$, the cosmological principle is
  therefore not sufficient to ensure that the expansion of an irreversible
  system follows the background expansion without violation of the second
  principle.

In an adiabatic system, the occurrence of an irreversible process and the
subsequent creation of entropy is possible through the dissipation of large
scale motions into microscopic
motions. This process, driven by virialization, is referred to as ``violent
relaxation'' within the context of structure formation in the Universe, in the
presence of non-collisional dark matter \citep{white1996}. When a significant portion
of the large-scale compression modes dissipates due to virialization, they can
no longer compensate the expansion occurring in cosmic voids, inevitably leading to
the conclusion that    $\langle \theta
\rangle \neq 3H$ (see bottom panel of
Fig.~\ref{fig:ir-reversible}). From numerical simulations \citep{haider2016}, we
can estimate that 50 \% of the dark matter mass is completely virialized in
haloes and 45 \% virialized in two directions in filaments \citep{eisenstein1997}. A realistic universe
has therefore virialized most of its mass and almost completely the initial
compression modes that appeared during the formation phase of the large scale structures.
Furthermore, it is crucial to emphasize that an
inhomogeneous system undergoing irreversible processes cannot, on average,
precisely mimic the behavior of the FLRW background. The background evolves akin
to a reversible system (constant entropy), while an inhomogeneous system subject to entropy
generation would essentially need to violate the second law of thermodynamics
with processes generating negative local entropy to
align its averaged entropy with the background entropy.
Additionally, the
process of dark matter virialization entails the acquisition of microscopic
stress through microscopic virialized motions. In this context, the dark matter
fluid ceases to be truly pressureless after virialization, resulting in a
non-zero stress tensor within the Jeans equations. This can be seen as a phase
transition to a bound state similar to the formation of a liquid but in the
context of statistical mechanics for non-ideal self-gravitating fluids \citep{tremblin2022}.
A central point to note is
that the equilibrium observed in large-scale structures is not a balance between
gravitational attraction and stresses induced by non-gravitational
interactions. Dark matter, by definition, is a tracer for dynamics along
time-like geodesics. Therefore, its virialization marks the dissipation of
large-scale dynamics along these geodesics into microscopic motions
that continue to follow geodesic trajectories in free-fall. This
fundamental aspect underscores why the virialization of large-scale structures
inevitably influences the cosmological redshift.

The presence of entropy creation, characterized by $\langle \vec{\nabla}\cdot
\vec{u_g} \rangle \neq 0$, necessarily implies the existence of discontinuities in
the three-velocity field. Remarkably, dark matter virialization occurs
subsequent to shell crossing, resulting in the formation of caustics, which are
essentially discontinuities in the three-velocity field. In the context of
entropy creation through violent relaxation, these discontinuities can be
likened to non-collisional accretion shocks (see \cite{parks2012,parks2017} for
non-collisional shocks in plasmas) and are associated to a microscopic stress
  tensor that is also discontinuous across the shocks/caustics.
The nature of such discontinuities within the framework of general
relativity may initially appear unclear. We
propose to characterize these discontinuities by utilizing exact solutions of
Einstein's field equations in spherical symmetry. These solutions are locally
comoving with the time-like geodesics and are known as generalized
Lemaître-Tolman-Bondi solutions, when accounting for a microscopic stress tensor and
virialization \citep{lasky2006}.

The metric for such solutions with a fluid characterized by an energy density $\rho$ and
pressure $\sigma$ is given by
\begin{equation}
ds^2 = -N(r,t)^2 dt^2 + \frac{1}{1+2E(r,t)}R^{\prime 2} dr^2 + R(r,t)^2 d\Omega^2,
\end{equation}
with $R^{\prime}$ the derivative with respect to $r$ and with a unit choice such
that $c=1$. $N(r,t)$ is the lapse function and $E(r,t)$ the local curvature. The
following Hamiltonian 
constraint equation can be derived from the ADM formalism \citep{lasky2006}
\begin{equation}
\frac{1}{2}u^2 = \left(\frac{GM}{R}+E\right)
\end{equation}
with  $u\equiv \dot{R}/N$, $\dot{R}$ the derivative with respect to $t$ and $M(r,t)=4\pi
\int_0^{R(r,t)}\rho \tilde{r}^2 d\tilde{r}$, with the following evolution
equations
\begin{eqnarray}
 \frac{\dot{E}}{N} &=& -\frac{1+2E}{\rho+\sigma}\frac{1}{R^\prime}\frac{\partial \sigma}{\partial r}u, \cr
 \frac{\dot{\rho} }{N}&=& -N(\rho+\sigma) \frac{1}{R^2 R^\prime}\frac{\partial
   (R^2 u)}{\partial r},\cr
 \frac{\dot{u}}{N}  &=& -\frac{GM}{R^2} - 4\pi G\sigma R -\frac{1}{R^\prime}
 \frac{1+2E}{\rho+\sigma}\frac{\partial \sigma}{\partial r}.
\end{eqnarray}
The Euler equation provides the following relation between the lapse $N$ and the
pressure $N^\prime/N = - \sigma^\prime/(\rho+\sigma)$. It is evident at this
stage that the local curvature $E(r,t)$ undergoes an evolution
reminiscent of the behavior of kinetic energy in classical Newtonian
hydrodynamics. This similarity suggests that the curvature can be linked to
microscopic dissipation through the presence of discontinuities. To establish
this connection, we proceed under the assumption that the fluid can be
characterized by an internal microscopic energy, denoted as $e$, such that the
total energy density $\rho$ can be expressed as $\rho=\rho_m+\rho_m e$, with
$\rho_m$ representing the mass density. It is important to emphasize that our
analysis holds validity for both collisional and non-collisional systems. The
microscopic internal energy can be defined for a non-colliosional system as the
microscopic stress tensor in the context of the Jeans equations. In the
non-collisional limit, we do not have simple closure relations that
establish a connection between the internal energy and the stress tensor with
other variables. However, it is worth noting that such a relation is not a
requisite element within our analysis. In the non-relativistic limit $E, e\ll
1$, $\sigma \ll \rho$. In this limit, and defining $\tau = 1/\rho_m$,
\begin{eqnarray}
  \dot{E} &=& -\frac{\tau}{R^\prime}\frac{\partial \sigma}{\partial r}u, \cr
  \dot{\tau} &=& \frac{\tau}{R^\prime}\frac{1}{R^2}\frac{\partial (R^2
    u)}{\partial r},\cr
  \dot{e} &=& - \frac{1}{\rho_m R^\prime}\frac{1}{R^2}\frac{\partial (R^2
    u)}{\partial r}\sigma,\cr
  \dot{u} &=& -\frac{GM}{R^2}-\frac{\tau}{ R^\prime}\frac{\partial
    \sigma}{\partial r}.
\end{eqnarray}
By employing the transformation $dm = \rho_m R^2 dr/R^\prime$, where $m$
represents the so-called "mass" variable, one can readily identify the
hydrodynamics equations in Lagrangian coordinates (see page 8 of
\cite{godlewski1996} or page 16 of \cite{despres2017}) with
an additional equation governing the evolution of curvature. From this system,
we can derive the following conservative equation that establishes a coupling
between curvature and internal energy:
\begin{equation}
\dot{E}+\dot{e} = -\frac{\partial}{\partial m}\left( \sigma u R^2\right).
\end{equation}
We assume now that a discontinuity is located at
$r_i$ with $R_i(t) = R(r_i,t)$, with the possibility for $u$ and $E$ to jump. By
using standard Rankine-Hugoniot relations,
\begin{equation}
\dot{m}_i \left(E_l+e_l-E_r-e_r\right) = \left(\sigma_l u_l-\sigma_r u_r \right) R^2_i,
\end{equation}
with $\dot{m}_i(t)$, the mass flow through the discontinuity. It is apparent
from this relation that the behavior of curvature bears a resemblance to that of
kinetic energy. The second law of thermodynamics, encompassing the dissipation
of large-scale motions and the concomitant generation of entropy, thus dictates
that the evolution of curvature follows the condition $\Delta E < 0$.
By applying the Hamiltonian constraint and using the continuity of $GM/R$
across the discontinuity, we can also derive the following relationship:
\begin{equation}
\frac{1}{2}\left(\dot{R}_l^2-\dot{R}_r^2 \right) = E_l-E_r.
\end{equation}
Under the assumption that the solution tends to become homogeneous on each side
of the interface, we arrive at the following jump condition concerning the local
volume expansion rate:
\begin{equation}
\frac{1}{18}\left(\theta_l^2-\theta_r^2 \right) = \frac{1}{R_i^2}\left(E_l-E_r\right).
\end{equation}
It is worth highlighting that these co-moving solutions can find utility in a
broader context not necessarily tied to gravitational dynamics. While we have
primarily considered particles following time-like geodesics, one can also
investigate co-moving solutions with non-geodesic motions in the limit as $G$
approaches zero. In this scenario, the solution becomes a spherical shock
discontinuity akin to the Sedov solution in classical hydrodynamics. Within this
context, $\theta_l = \vec{\nabla}\cdot \vec{u_g}$ remains approximately constant
within the volume undergoing expansion, while $\theta_r = 0$ outside of it.
These shocks dissipate large-scale motions into
microscopic motions, accompanied by entropy creation and a distinctive jump in
the local curvature, the evolution of which conforms to $\Delta E<0$. Commencing
with a homogeneous system characterized by zero curvature and transitioning to a
locally comoving space-time with time-like geodesics, we find that virialization
leads to the generation of local negative curvature. Although the connection
between virialization and negative curvature has been discussed in existing
literature \citep{roukema2013,roukema2018}, our contribution lies in linking
the curvature production by virialization to the second law of
thermodynamics. This implies that the averaged space-time curvature of a
co-moving solution that traces the motion of time-like geodesics, cannot be zero
in the presence of irreversible processes.

\begin{figure}[!ht]
\centering
\includegraphics[width=0.8\linewidth]{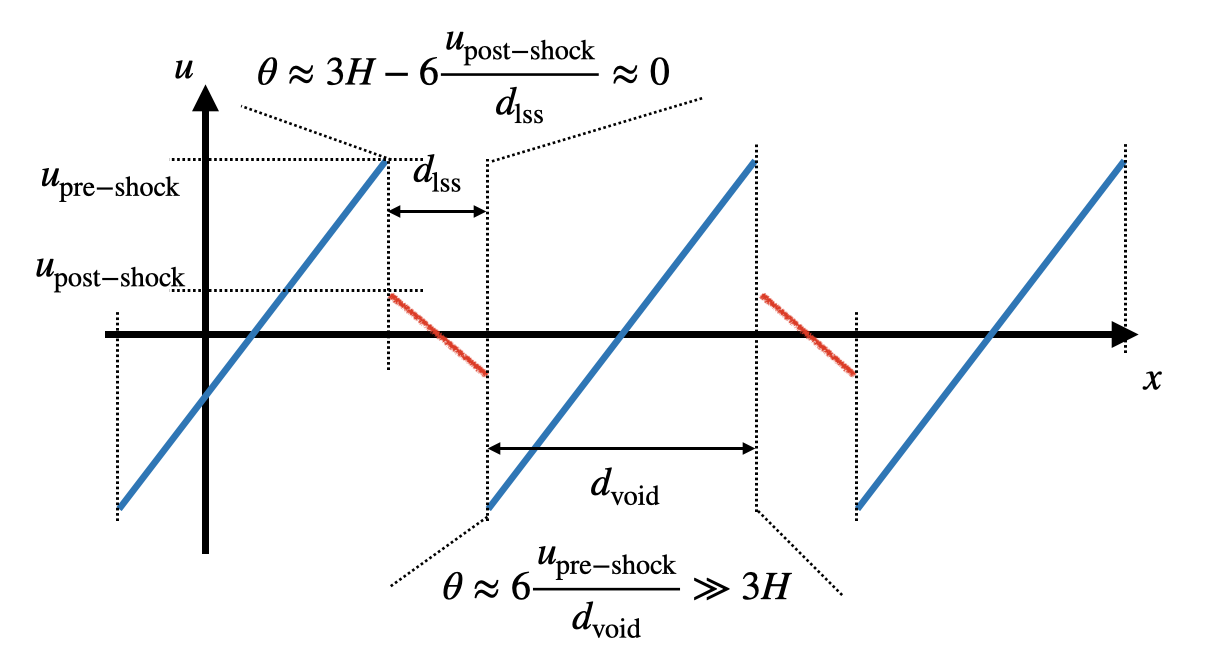}
\includegraphics[width=0.8\linewidth]{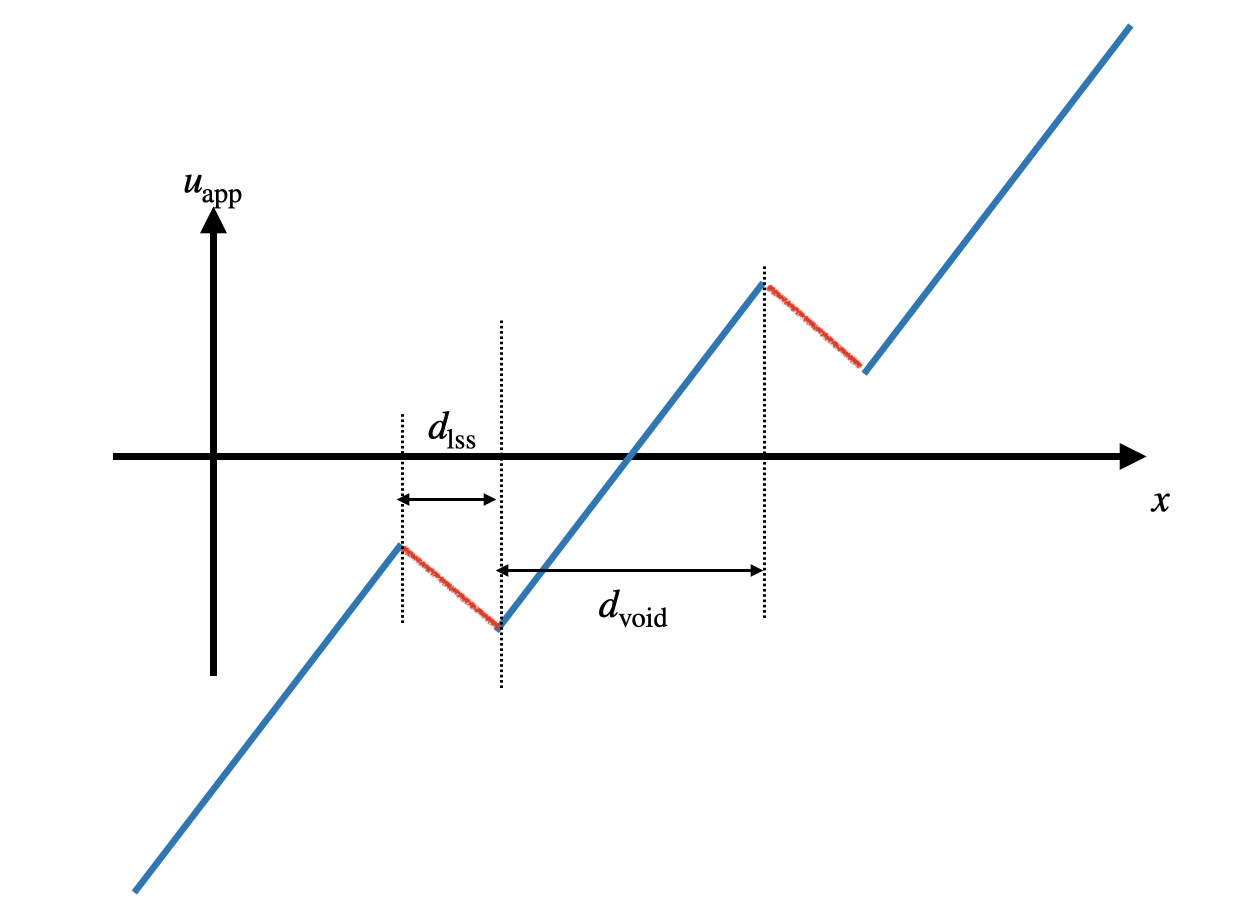}
\caption{Top: Illustration of a discontinuous velocity profile that impacts the
  redshift-distance relation. Bottom: equivalent apparent velocity profile
  by integration of the gradients along the line-of-sight.}\label{fig:distance-redshift}
\end{figure}

\section{Link with observations and quantitative estimates}\label{sect:obs}

\cite{adame2023,adame2024} present the release of the first science data from the DESI (Dark Energy
Spectroscopic Instrument) project, which includes data from commissioning and
Survey Validation phases conducted between December 2020 and June
2021. The authors highlight that the DESI collaboration has successfully validated its
survey design and observing strategy, ensuring the data's quality and
completeness and that future releases will facilitate
detailed studies of the universe's large-scale structure, providing critical
insights into dark energy and the fundamental physics governing the universe's
evolution. Following the initial data release, the DESI collaboration has
unveiled its first-year analysis results of baryon 
acoustic oscillations (BAO) based on extensive observations of galaxies,
quasars, and the Lyman-$\alpha$ forest \citep{adame2024b}. While the DESI data alone
align with the $\Lambda$CDM 
model, the results deviate when generalized to the $\omega_0 \omega_a$CDM
model. The DESI data combined
with cosmic microwave background (CMB) and supernova data, suggests a rejection
of the  $\Lambda$CDM model in favor of the $\omega_0 \omega_a$CDM model with
significant confidence levels between 2 and 4$\sigma$ \citep{adame2024b}. The combined data
indicate a preference for $\omega_0 > -1$ and $\omega_a
< 0$, where $\omega(a) = \omega_0 + \omega_a(1 - a)$ represents the equation-of-state
parameter of dark energy with $a(t)$ the scale factor of the FLRW
cosmology. If confirmed, these findings could have profound implications for our
understanding of the Universe's dynamics. In that context, \cite{tada2024}
suggest that the nature of dark energy might be explained
by a quintessential scalar field, although this field could potentially become a
phantom in the past. They conclude that the DESI data challenges the
$\Lambda$CDM model and supports a dynamic dark energy scenario, 
thereby contributing to our understanding of cosmic acceleration and the
potential need for new physics beyond the standard cosmological model.

However, due to the covariance of General Relativity, the current approach in the
standard cosmological model can be reformulated as follows: the choice of an
expanding background coordinate system should be arbitrary and introducing a
cosmological constant to drive the dynamics of the coordinate system has no
physical meaning other than selecting an expanding coordinate system with
$a_\mathrm{eff}(t)$, $H_\mathrm{eff}(t)$, and three velocities $\vec{u_\mathrm{eff,g}}$
for time-like geodesics such that $\langle \theta \rangle \approx 3H_\mathrm{eff}$
and $\langle \vec{\nabla} \cdot \vec{u_\mathrm{eff,g}} \rangle \approx 0$. 
This choice compensates in the coordinate system the averaged expansion rate
along time-like geodesics present in the three-velocity field. If one prefers to
use the FLRW solution of a fictive homogeneous universe
without a cosmological constant, where the Hubble expansion rate is $H(t)$ (with
$H^2=8\pi G\bar{\rho}/3$),
and the three velocities for time-like geodesics are $\vec{u_g}$, then
$\langle \theta \rangle = 3H +\langle \vec{\nabla} \cdot \vec{u_g}\rangle$ and $\langle
\vec{\nabla} \cdot \vec{u_g} \rangle \neq 0$ as soon as virialization and
entropy production occur.
We highlight that virialization during structure formation is likely to replace
dark energy because it precisely mimics the counter-intuitive anti-gravity
effect needed for dark energy. The dissipation of large-scale compression modes
by virialization stabilizes large-scale gravitational collapse (anti-gravity)
while remaining a purely gravitational effect, as virialized dark matter
continues to be in free-fall.

In this effective coordinate system, we can use
the second principle of thermodynamics to infer an effective evolution equation
that takes into account entropy production,
\begin{eqnarray}
 \frac{ d(\log V_\mathrm{eff})}{dt} = \frac{d(\log
   V_\mathrm{rev})}{dt}+\frac{1}{Z k_b}\frac{ds}{dt},\cr
 H_\mathrm{eff}(t) = H(t) +\frac{1}{3 Z k_b}\frac{ds}{dt},
\end{eqnarray}
which can be rewritten as,
\begin{eqnarray}
  H_\mathrm{eff}(t)^2 = \frac{8\pi G \bar{\rho}}{3} + \frac{\Lambda_\mathrm{eff}(t) c^2}{3},\cr
\Lambda_\mathrm{eff}(t) = \frac{2H}{c^2 Z k_b}\frac{ds}{dt} + \frac{1}{3 c^2 Z^2
  k_b^2}\left(\frac{ds}{dt}\right)^2.
\end{eqnarray}
In this context, the effective cosmological constant appears to vary with time,
as suggested by the early DESI data release. It is important to note, however,
that this does not imply the existence of new physics; rather, the effective
cosmological constant serves as a proxy to account for the impact of entropy
production on the averaged expansion rate along time-like geodesics. Entropy
production is expected to begin during structure formation at the first shell
crossing, peak during the virialization of structures, and gradually decline to
the present day. This behavior could simulate the need for $\omega(a) < -1$ in
the past and $\omega(a) > -1$ in the present.
As noted in \cite{tada2024}, The increase of $\omega(a)$ is correlated with the
decrease of the Hubble constant, which is the opposite direction to solve the
Hubble tension. The Hubble Tension refers to the significant discrepancy between
two different methods of measuring the Hubble constant. The first method
involves local measurements, such 
as observing the distances and redshifts of nearby galaxies. The second method
involves the Cosmic Microwave Background (CMB) and large-scale structure data,
which rely on $\Lambda$CDM model and measurements from the
early Universe \citep{kamionkowski2023}. The local measurements consistently
yield a higher value for the 
Hubble constant (around 73 km/s/Mpc) compared to the CMB-based measurements
(around 67
km/s/Mpc). Although the decline phase of entropy production after the
virialization of structures cannot explain the Hubble tension, the initial
increase phase after the first shell crossing could play an important role in
resolving the current tension. However, precise estimates of
the evolution of $\Lambda_\mathrm{eff}(t)$ require dedicated numerical
simulations to accurately constrain the evolution of entropy production during
structure formation and subsequent evolution.

We can, however, use existing numerical simulations to provide a more
quantitative order-of-magnitude estimate. Let us consider the illustrative
example depicted in the top panel of Fig.~\ref{fig:distance-redshift}. In this
scenario, we
assume that within the bound and virialized large-scale structures with a
characteristic size of $d_\mathrm{lss}$, the local volume expansion rate,
denoted as $\theta$, is approximately negligible ($\theta\approx
0$). Conversely, in the expanding cosmic voids with a typical size of
$d_\mathrm{void}$, $\theta$ is substantially higher, with $\theta \approx
6u_\mathrm{pre-shock}/d_\mathrm{void}\gg 3H$. The bottom panel of
Fig.~\ref{fig:distance-redshift} shows an
equivalent apparent velocity profile by integrating the gradients along the
line-of-sight. Both profiles in Fig.~\ref{fig:distance-redshift} provide the
same integrated redshift using eq.~\ref{eq:cov-redshift2} and illustrate how
peculiar velocities resulting from structure formation can mimic a Hubble flow.
Without discontinuities in Fig.~\ref{fig:distance-redshift}, the
  velocity field divergence in the structures
and in the voids would exactly compensates such that
$\langle \vec{\nabla}\cdot \vec{u_g} \rangle =0$. However, this is not the case with
discontinuities: in the extreme case for which $ \vec{\nabla}\cdot \vec{u_g} $ is
strictly positive everywhere (i.e. with expanding motions in the lss regions in
Fig.~\ref{fig:distance-redshift}), $\langle \vec{\nabla}\cdot \vec{u_g} \rangle
$ is clearly strictly positive on average. 
In a realistic configuration for the large scale structures with $\theta\approx
0$, the dominant contributor to the redshift
arises from the expansion occurring within the voids, accounting for a fraction
of $d_\mathrm{void}/(d_\mathrm{lss}+d_\mathrm{void})$. This expansion in the
voids cannot be counterbalanced by compression within the structures, as it has
dissipated due to violent relaxation and virialization processes. Assuming
typical structure sizes of approximately $d_\mathrm{lss}\approx$ 3 Mpc and void
sizes around $d_\mathrm{void}\approx$ 30 Mpc, the average expansion rate,
represented by $\langle \theta \rangle /3$, attains values on the order of 70
km/s/Mpc with pre-shock velocities of roughly
$u_\mathrm{pre-shock}\approx 1200$ km/s. This estimation falls within the same order of
magnitude as the infall velocities observed in N-body numerical simulations
\citep{zu2013}.

The forthcoming generation of high-resolution cosmological simulations on
exascale supercomputers will be essential in advancing our understanding of
entropy production through caustics (i.e., velocity discontinuities or
non-collisional shocks). These simulations will enable more precise estimates of
the averaged local expansion rate along time-like geodesics. Such estimates are
crucial for interpreting dynamic dark energy scenarios, where dark energy could
be replaced by entropy production during structure formation. These insights
will be essential for analyzing future data releases from the DESI collaboration
and for interpreting observations from the Euclid satellite
\citep{amendola2018}. By combining
Euclid's weak lensing data, redshift surveys (including baryon acoustic
oscillations, redshift distortions, and the full power spectrum shape), and CMB
data from Planck, we can constrain dynamic dark energy scenarios more
effectively.

\section{Conclusions}\label{sect:con}

In conclusion, this paper highlights the intricate relationship between the
cosmological redshift and entropy production due to structure formation. We
emphasize the need of the covariant calculation of the cosmological redshift,
which accounts for peculiar motions with a local expansion rate along time-like
geodesics.

Our analysis suggests that virialization and the subsequent entropy production
during structure formation play a crucial role in this dynamic. The time-varying
nature of the effective cosmological constant, as indicated by recent DESI data,
may be a manifestation of these processes rather than new physics. This
perspective necessitates a reassessment of dark energy models, proposing that
what is currently attributed to dark energy might instead be the result of
entropy production during structure formation.

Future high-resolution cosmological simulations on exascale supercomputers will
be vital for refining our understanding of these phenomena. These simulations
will help produce more accurate estimates of the averaged local expansion rate
along time-like geodesics, providing essential insights for interpreting dynamic
dark energy scenarios and forthcoming observational data from DESI and the
Euclid satellite. This research opens new avenues for understanding the
fundamental forces shaping our universe, emphasizing the need for continuous
advancements in computational and observational cosmology.

\begin{acknowledgements}
  We thank Quentin Vigneron, Thomas Guillet, and Pascal Wang for all the discussions that lead to
this paper. We also thank David Elbaz and Matthias Gonz\'alez for helpful
comments on this manuscript.
\end{acknowledgements}

%
%

\bibliography{main}

\begin{thebibliography}{30}
\expandafter\ifx\csname natexlab\endcsname\relax\def\natexlab#1{#1}\fi

\bibitem[{{Amendola} {et~al.}(2018){Amendola}, {Appleby}, {Avgoustidis},
  {Bacon}, {Baker}, {Baldi}, {Bartolo}, {Blanchard}, {Bonvin}, {Borgani},
  {Branchini}, {Burrage}, {Camera}, {Carbone}, {Casarini}, {Cropper}, {de
  Rham}, {Dietrich}, {Di Porto}, {Durrer}, {Ealet}, {Ferreira}, {Finelli},
  {Garc{\'\i}a-Bellido}, {Giannantonio}, {Guzzo}, {Heavens}, {Heisenberg},
  {Heymans}, {Hoekstra}, {Hollenstein}, {Holmes}, {Hwang}, {Jahnke},
  {Kitching}, {Koivisto}, {Kunz}, {La Vacca}, {Linder}, {March}, {Marra},
  {Martins}, {Majerotto}, {Markovic}, {Marsh}, {Marulli}, {Massey}, {Mellier},
  {Montanari}, {Mota}, {Nunes}, {Percival}, {Pettorino}, {Porciani},
  {Quercellini}, {Read}, {Rinaldi}, {Sapone}, {Sawicki}, {Scaramella},
  {Skordis}, {Simpson}, {Taylor}, {Thomas}, {Trotta}, {Verde}, {Vernizzi},
  {Vollmer}, {Wang}, {Weller}, \& {Zlosnik}}]{amendola2018}
{Amendola}, L., {Appleby}, S., {Avgoustidis}, A., {et~al.} 2018, Living Reviews
  in Relativity, 21, 2

\bibitem[{{Buchert}(2008)}]{buchert2008}
{Buchert}, T. 2008, General Relativity and Gravitation, 40, 467

\bibitem[{{Buchert} {et~al.}(2015){Buchert}, {Carfora}, {Ellis}, {Kolb},
  {MacCallum}, {Ostrowski}, {R{\"a}s{\"a}nen}, {Roukema}, {Andersson}, {Coley},
  \& {Wiltshire}}]{buchert2015}
{Buchert}, T., {Carfora}, M., {Ellis}, G.~F.~R., {et~al.} 2015, Classical and
  Quantum Gravity, 32, 215021

\bibitem[{{Buchert} \& {Ehlers}(1997)}]{buchert1997}
{Buchert}, T. \& {Ehlers}, J. 1997, A\&A, 320, 1

\bibitem[{{Bunn} \& {Hogg}(2009)}]{bunn2009}
{Bunn}, E.~F. \& {Hogg}, D.~W. 2009, American Journal of Physics, 77, 688

\bibitem[{{Chodorowski}(2007)}]{chodorowski2007}
{Chodorowski}, M.~J. 2007, MNRAS, 378, 239

\bibitem[{{DESI Collaboration} {et~al.}(2023){DESI Collaboration}, {Adame},
  {Aguilar}, {Ahlen}, {Alam}, {Aldering}, {Alexander}, {Alfarsy}, {Allende
  Prieto}, {Alvarez}, {Alves}, {Anand}, {Andrade-Oliveira}, {Armengaud},
  {Asorey}, {Avila}, {Aviles}, {Bailey}, {Balaguera-Antol{\'\i}nez},
  {Ballester}, {Baltay}, {Bault}, {Bautista}, {Behera}, {Beltran}, {BenZvi},
  {Beraldo e Silva}, {Bermejo-Climent}, {Berti}, {Besuner}, {Beutler},
  {Bianchi}, {Blake}, {Blum}, {Bolton}, {Brieden}, {Brodzeller}, {Brooks},
  {Brown}, {Buckley-Geer}, {Burtin}, {Cabayol-Garcia}, {Cai}, {Canning},
  {Cardiel-Sas}, {Carnero Rosell}, {Castander}, {Cervantes-Cota}, {Chabanier},
  {Chaussidon}, {Chaves-Montero}, {Chen}, {Chuang}, {Claybaugh}, {Cole},
  {Cooper}, {Cuceu}, {Davis}, {Dawson}, {de Belsunce}, {de la Cruz}, {de la
  Macorra}, {de Mattia}, {Demina}, {Demirbozan}, {DeRose}, {Dey}, {Dey},
  {Dhungana}, {Ding}, {Ding}, {Doel}, {Doshi}, {Douglass}, {Edge},
  {Eftekharzadeh}, {Eisenstein}, {Elliott}, {Escoffier}, {Fagrelius}, {Fan},
  {Fanning}, {Fawcett}, {Ferraro}, {Ereza}, {Flaugher}, {Font-Ribera},
  {Forero-S{\'a}nchez}, {Forero-Romero}, {Frenk}, {G{\"a}nsicke},
  {Garc{\'\i}a}, {Garc{\'\i}a-Bellido}, {Garcia-Quintero}, {Garrison},
  {Gil-Mar{\'\i}n}, {Golden-Marx}, {Gontcho}, {Gonzalez-Morales},
  {Gonzalez-Perez}, {Gordon}, {Graur}, {Green}, {Gruen}, {Guy}, {Hadzhiyska},
  {Hahn}, {Han}, {Hanif}, {Herrera-Alcantar}, {Honscheid}, {Hou}, {Howlett},
  {Huterer}, {Ir{\v{s}}i{\v{c}}}, {Ishak}, {Jacques}, {Jana}, {Jiang},
  {Jimenez}, {Jing}, {Joudaki}, {Jullo}, {Juneau}, {Kizhuprakkat},
  {Kara{\c{c}}ayl{\i}}, {Karim}, {Kehoe}, {Kent}, {Khederlarian}, {Kim},
  {Kirkby}, {Kisner}, {Kitaura}, {Kneib}, {Koposov}, {Kov{\'a}cs}, {Kremin},
  {Krolewski}, {L'Huillier}, {Lambert}, {Lamman}, {Lan}, {Landriau}, {Lang},
  {Lange}, {Lasker}, {Le Guillou}, {Leauthaud}, {Levi}, {Li}, {Linder},
  {Lyons}, {Magneville}, {Manera}, {Manser}, {Margala}, {Martini}, {McDonald},
  {Medina}, {Medina-Varela}, {Meisner}, {Mena-Fern{\'a}ndez}, {Meneses-Rizo},
  {Mezcua}, {Miquel}, {Montero-Camacho}, {Moon}, {Moore}, {Moustakas},
  {Mueller}, {Mundet}, {Mu{\~n}oz-Guti{\'e}rrez}, {Myers}, {Nadathur},
  {Napolitano}, {Neveux}, {Newman}, {Nie}, {Nikutta}, {Niz}, {Norberg},
  {Noriega}, {Paillas}, {Palanque-Delabrouille}, {Palmese}, {Zhiwei},
  {Parkinson}, {Penmetsa}, {Percival}, {P{\'e}rez-Fern{\'a}ndez},
  {P{\'e}rez-R{\`a}fols}, {Pieri}, {Poppett}, {Porredon}, {Pothier}, {Prada},
  {Pucha}, {Raichoor}, {Ram{\'\i}rez-P{\'e}rez}, {Ramirez-Solano},
  {Rashkovetskyi}, {Ravoux}, {Rocher}, {Rockosi}, {Ross}, {Rossi}, {Ruggeri},
  {Ruhlmann-Kleider}, {Sabiu}, {Said}, {Saintonge}, {Samushia}, {Sanchez},
  {Saulder}, {Schaan}, {Schlafly}, {Schlegel}, {Scholte}, {Schubnell}, {Seo},
  {Shafieloo}, {Sharples}, {Sheu}, {Silber}, {Sinigaglia}, {Siudek}, {Slepian},
  {Smith}, {Sprayberry}, {Stephey}, {Su{\'a}rez-P{\'e}rez}, {Sun}, {Tan},
  {Tarl{\'e}}, {Tojeiro}, {Ure{\~n}a-L{\'o}pez}, {Vaisakh}, {Valcin}, {Valdes},
  {Valluri}, {Vargas-Maga{\~n}a}, {Variu}, {Verde}, {Walther}, {Wang}, {Wang},
  {Weaver}, {Weaverdyck}, {Wechsler}, {White}, {Xie}, {Yang}, {Y{\`e}che},
  {Yu}, {Yuan}, {Zhang}, {Zhang}, {Zhao}, {Zheng}, {Zhou}, {Zhou}, {Zou},
  {Zou}, \& {Zu}}]{adame2023}
{DESI Collaboration}, {Adame}, A.~G., {Aguilar}, J., {et~al.} 2023, arXiv
  e-prints, arXiv:2306.06308

\bibitem[{{DESI Collaboration} {et~al.}(2024{\natexlab{a}}){DESI
  Collaboration}, {Adame}, {Aguilar}, {Ahlen}, {Alam}, {Aldering}, {Alexander},
  {Alfarsy}, {Allende Prieto}, {Alvarez}, {Alves}, {Anand}, {Andrade-Oliveira},
  {Armengaud}, {Asorey}, {Avila}, {Aviles}, {Bailey},
  {Balaguera-Antol{\'\i}nez}, {Ballester}, {Baltay}, {Bault}, {Bautista},
  {Behera}, {Beltran}, {BenZvi}, {Beraldo e Silva}, {Bermejo-Climent}, {Berti},
  {Besuner}, {Beutler}, {Bianchi}, {Blake}, {Blum}, {Bolton}, {Brieden},
  {Brodzeller}, {Brooks}, {Brown}, {Buckley-Geer}, {Burtin}, {Cabayol-Garcia},
  {Cai}, {Canning}, {Cardiel-Sas}, {Carnero Rosell}, {Castander},
  {Cervantes-Cota}, {Chabanier}, {Chaussidon}, {Chaves-Montero}, {Chen},
  {Chen}, {Chuang}, {Claybaugh}, {Cole}, {Cooper}, {Cuceu}, {Davis}, {Dawson},
  {de Belsunce}, {de la Cruz}, {de la Macorra}, {de Mattia}, {Demina},
  {Demirbozan}, {DeRose}, {Dey}, {Dey}, {Dhungana}, {Ding}, {Ding}, {Doel},
  {Doshi}, {Douglass}, {Edge}, {Eftekharzadeh}, {Eisenstein}, {Elliott},
  {Escoffier}, {Fagrelius}, {Fan}, {Fanning}, {Fawcett}, {Ferraro}, {Ereza},
  {Flaugher}, {Font-Ribera}, {Forero-S{\'a}nchez}, {Forero-Romero}, {Frenk},
  {G{\"a}nsicke}, {Garc{\'\i}a}, {Garc{\'\i}a-Bellido}, {Garcia-Quintero},
  {Garrison}, {Gil-Mar{\'\i}n}, {Golden-Marx}, {Gontcho A Gontcho},
  {Gonzalez-Morales}, {Gonzalez-Perez}, {Gordon}, {Graur}, {Green}, {Gruen},
  {Guy}, {Hadzhiyska}, {Hahn}, {Han}, {Hanif}, {Herrera-Alcantar}, {Honscheid},
  {Hou}, {Howlett}, {Huterer}, {Ir{\v{s}}i{\v{c}}}, {Ishak}, {Jana}, {Jiang},
  {Jimenez}, {Jing}, {Joudaki}, {Jullo}, {Joyce}, {Juneau}, {Kizhuprakkat},
  {Kara{\c{c}}ayl{\i}}, {Karim}, {Kehoe}, {Kent}, {Khederlarian}, {Kim},
  {Kirkby}, {Kisner}, {Kitaura}, {Kneib}, {Koposov}, {Kov{\'a}cs}, {Kremin},
  {Krolewski}, {L'Huillier}, {Lahav}, {Lambert}, {Lamman}, {Lan}, {Landriau},
  {Lang}, {Lange}, {Lasker}, {Le Guillou}, {Leauthaud}, {Levi}, {Li}, {Linder},
  {Lyons}, {Magneville}, {Manera}, {Manser}, {Margala}, {Martini}, {McDonald},
  {Medina}, {Medina-Varela}, {Meisner}, {Mena-Fern{\'a}ndez}, {Meneses-Rizo},
  {Mezcua}, {Miquel}, {Montero-Camacho}, {Moon}, {Moore}, {Moustakas},
  {Mueller}, {Mundet}, {Mu{\~n}oz-Guti{\'e}rrez}, {Myers}, {Nadathur},
  {Napolitano}, {Neveux}, {Newman}, {Nie}, {Niz}, {Norberg}, {Noriega},
  {Paillas}, {Palanque-Delabrouille}, {Palmese}, {Zhiwei}, {Parkinson},
  {Penmetsa}, {Percival}, {P{\'e}rez-Fern{\'a}ndez}, {P{\'e}rez-R{\`a}fols},
  {Pieri}, {Poppett}, {Porredon}, {Prada}, {Pucha}, {Raichoor},
  {Ram{\'\i}rez-P{\'e}rez}, {Ramirez-Solano}, {Rashkovetskyi}, {Ravoux},
  {Rocher}, {Rockosi}, {Ross}, {Rossi}, {Ruggeri}, {Ruhlmann-Kleider}, {Sabiu},
  {Said}, {Saintonge}, {Samushia}, {Sanchez}, {Saulder}, {Schaan}, {Schlafly},
  {Schlegel}, {Scholte}, {Schubnell}, {Seo}, {Shafieloo}, {Sharples}, {Sheu},
  {Silber}, {Sinigaglia}, {Siudek}, {Slepian}, {Smith}, {Sprayberry},
  {Stephey}, {Su{\'a}rez-P{\'e}rez}, {Sun}, {Tan}, {Tarl{\'e}}, {Tojeiro},
  {Ure{\~n}a-L{\'o}pez}, {Vaisakh}, {Valcin}, {Valdes}, {Valluri},
  {Vargas-Maga{\~n}a}, {Variu}, {Verde}, {Walther}, {Wang}, {Wang}, {Weaver},
  {Weaverdyck}, {Wechsler}, {White}, {Xie}, {Yang}, {Y{\`e}che}, {Yu}, {Yuan},
  {Zhang}, {Zhang}, {Zhao}, {Zheng}, {Zhou}, {Zhou}, {Zou}, {Zou}, {Zu}, \&
  {DESI Collaboration}}]{adame2024}
{DESI Collaboration}, {Adame}, A.~G., {Aguilar}, J., {et~al.}
  2024{\natexlab{a}}, \aj, 167, 62

\bibitem[{{DESI Collaboration} {et~al.}(2024{\natexlab{b}}){DESI
  Collaboration}, {Adame}, {Aguilar}, {Ahlen}, {Alam}, {Alexander}, {Alvarez},
  {Alves}, {Anand}, {Andrade}, {Armengaud}, {Avila}, {Aviles}, {Awan},
  {Bahr-Kalus}, {Bailey}, {Baltay}, {Bault}, {Behera}, {BenZvi}, {Bera},
  {Beutler}, {Bianchi}, {Blake}, {Blum}, {Brieden}, {Brodzeller}, {Brooks},
  {Buckley-Geer}, {Burtin}, {Calderon}, {Canning}, {Carnero Rosell},
  {Cereskaite}, {Cervantes-Cota}, {Chabanier}, {Chaussidon}, {Chaves-Montero},
  {Chen}, {Chen}, {Claybaugh}, {Cole}, {Cuceu}, {Davis}, {Dawson}, {de la
  Macorra}, {de Mattia}, {Deiosso}, {Dey}, {Dey}, {Ding}, {Doel}, {Edelstein},
  {Eftekharzadeh}, {Eisenstein}, {Elliott}, {Fagrelius}, {Fanning}, {Ferraro},
  {Ereza}, {Findlay}, {Flaugher}, {Font-Ribera}, {Forero-S{\'a}nchez},
  {Forero-Romero}, {Frenk}, {Garcia-Quintero}, {Gazta{\~n}aga},
  {Gil-Mar{\'\i}n}, {Gontcho}, {Gonzalez-Morales}, {Gonzalez-Perez}, {Gordon},
  {Green}, {Gruen}, {Gsponer}, {Gutierrez}, {Guy}, {Hadzhiyska}, {Hahn},
  {Hanif}, {Herrera-Alcantar}, {Honscheid}, {Howlett}, {Huterer},
  {Ir{\v{s}}i{\v{c}}}, {Ishak}, {Juneau}, {Kara{\c{c}}ayl{\i}}, {Kehoe},
  {Kent}, {Kirkby}, {Kremin}, {Krolewski}, {Lai}, {Lan}, {Landriau}, {Lang},
  {Lasker}, {Le Goff}, {Le Guillou}, {Leauthaud}, {Levi}, {Li}, {Linder},
  {Lodha}, {Magneville}, {Manera}, {Margala}, {Martini}, {Maus}, {McDonald},
  {Medina-Varela}, {Meisner}, {Mena-Fern{\'a}ndez}, {Miquel}, {Moon}, {Moore},
  {Moustakas}, {Mudur}, {Mueller}, {Mu{\~n}oz-Guti{\'e}rrez}, {Myers},
  {Nadathur}, {Napolitano}, {Neveux}, {Newman}, {Nguyen}, {Nie}, {Niz},
  {Noriega}, {Padmanabhan}, {Paillas}, {Palanque-Delabrouille}, {Pan},
  {Penmetsa}, {Percival}, {Pieri}, {Pinon}, {Poppett}, {Porredon}, {Prada},
  {P{\'e}rez-Fern{\'a}ndez}, {P{\'e}rez-R{\`a}fols}, {Rabinowitz}, {Raichoor},
  {Ram{\'\i}rez-P{\'e}rez}, {Ramirez-Solano}, {Ravoux}, {Rashkovetskyi},
  {Rezaie}, {Rich}, {Rocher}, {Rockosi}, {Roe}, {Rosado-Marin}, {Ross},
  {Rossi}, {Ruggeri}, {Ruhlmann-Kleider}, {Samushia}, {Sanchez}, {Saulder},
  {Schlafly}, {Schlegel}, {Schubnell}, {Seo}, {Shafieloo}, {Sharples},
  {Silber}, {Slosar}, {Smith}, {Sprayberry}, {Tan}, {Tarl{\'e}}, {Taylor},
  {Trusov}, {Ure{\~n}a-L{\'o}pez}, {Vaisakh}, {Valcin}, {Valdes},
  {Vargas-Maga{\~n}a}, {Verde}, {Walther}, {Wang}, {Wang}, {Weaver},
  {Weaverdyck}, {Wechsler}, {Weinberg}, {White}, {Yu}, {Yu}, {Yuan},
  {Y{\`e}che}, {Zaborowski}, {Zarrouk}, {Zhang}, {Zhao}, {Zhao}, {Zhou},
  {Zhuang}, \& {Zou}}]{adame2024b}
{DESI Collaboration}, {Adame}, A.~G., {Aguilar}, J., {et~al.}
  2024{\natexlab{b}}, arXiv e-prints, arXiv:2404.03002

\bibitem[{Despr{\'e}s(2017)}]{despres2017}
Despr{\'e}s, B. 2017, Numerical Methods for Eulerian and Lagrangian
  Conservation Laws, Frontiers in Mathematics (Springer International
  Publishing)

\bibitem[{{Eisenstein} {et~al.}(1997){Eisenstein}, {Loeb}, \&
  {Turner}}]{eisenstein1997}
{Eisenstein}, D.~J., {Loeb}, A., \& {Turner}, E.~L. 1997, ApJ, 475, 421

\bibitem[{Godlewski \& Raviart(1996)}]{godlewski1996}
Godlewski, E. \& Raviart, P. 1996, Numerical Approximation of Hyperbolic
  Systems of Conservation Laws, Applied Mathematical Sciences No. 118
  (Springer)

\bibitem[{{Haider} {et~al.}(2016){Haider}, {Steinhauser}, {Vogelsberger},
  {Genel}, {Springel}, {Torrey}, \& {Hernquist}}]{haider2016}
{Haider}, M., {Steinhauser}, D., {Vogelsberger}, M., {et~al.} 2016, MNRAS, 457,
  3024

\bibitem[{{Ishibashi} \& {Wald}(2006)}]{ishibashi2006}
{Ishibashi}, A. \& {Wald}, R.~M. 2006, Classical and Quantum Gravity, 23, 235

\bibitem[{Kamionkowski \& Riess(2023)}]{kamionkowski2023}
Kamionkowski, M. \& Riess, A.~G. 2023, Annual Review of Nuclear and Particle
  Science, 73, 153

\bibitem[{{Lasky} \& {Lun}(2006)}]{lasky2006}
{Lasky}, P.~D. \& {Lun}, A.~W.~C. 2006, PRD, 74, 084013

\bibitem[{{Parks} {et~al.}(2017){Parks}, {Lee}, {Fu}, {Lin}, {Liu}, \&
  {Yang}}]{parks2017}
{Parks}, G.~K., {Lee}, E., {Fu}, S.~Y., {et~al.} 2017, Reviews of Modern Plasma
  Physics, 1, 1

\bibitem[{{Parks} {et~al.}(2012){Parks}, {Lee}, {McCarthy}, {Goldstein}, {Fu},
  {Cao}, {Canu}, {Lin}, {Wilber}, {Dandouras}, {R{\'e}me}, \&
  {Fazakerley}}]{parks2012}
{Parks}, G.~K., {Lee}, E., {McCarthy}, M., {et~al.} 2012, PRL, 108, 061102

\bibitem[{Peacock(1998)}]{peacock1998}
Peacock, J.~A. 1998, Cosmological Physics (Cambridge University Press)

\bibitem[{Peacock(2008)}]{peacock2008}
Peacock, J.~A. 2008, A diatribe on expanding space

\bibitem[{Peebles(1993)}]{peebles1993}
Peebles, P. 1993, Principles of Physical Cosmology, Princeton Series in Physics
  (Princeton University Press)

\bibitem[{{R{\"a}s{\"a}nen}(2009)}]{rasanen2009}
{R{\"a}s{\"a}nen}, S. 2009, JCAP, 2009, 011

\bibitem[{Rasanen(2010)}]{rasanen2010}
Rasanen, S. 2010, Journal of Cosmology and Astroparticle Physics, 2010

\bibitem[{{Roukema}(2018)}]{roukema2018}
{Roukema}, B.~F. 2018, A\&A, 610, A51

\bibitem[{{Roukema} {et~al.}(2013){Roukema}, {Ostrowski}, \&
  {Buchert}}]{roukema2013}
{Roukema}, B.~F., {Ostrowski}, J.~J., \& {Buchert}, T. 2013, JCAP, 2013, 043

\bibitem[{{Tada} \& {Terada}(2024)}]{tada2024}
{Tada}, Y. \& {Terada}, T. 2024, \prd, 109, L121305

\bibitem[{{Tremblin} {et~al.}(2022){Tremblin}, {Chabrier}, {Padioleau}, \&
  {Daley-Yates}}]{tremblin2022}
{Tremblin}, P., {Chabrier}, G., {Padioleau}, T., \& {Daley-Yates}, S. 2022,
  A\&A, 659, A108

\bibitem[{{White}(1996)}]{white1996}
{White}, S.~D.~M. 1996, in Gravitational dynamics, ed. O.~{Lahav},
  E.~{Terlevich}, \& R.~J. {Terlevich}, 121

\bibitem[{Whiting(2004)}]{whiting2004}
Whiting, A.~B. 2004, arXiv: Astrophysics

\bibitem[{{Zu} \& {Weinberg}(2013)}]{zu2013}
{Zu}, Y. \& {Weinberg}, D.~H. 2013, MNRAS, 431, 3319

\end{thebibliography}
\bibliographystyle{aa}

\begin{appendix} 

\end{appendix}

\end{document}